\documentclass[twocolumn,10pt]{article}
\usepackage[height=9in,width=7in]{geometry}
\usepackage{times}
\usepackage[T1]{fontenc}
\usepackage{amsmath,amsfonts,amsthm}
\usepackage{bm}
\usepackage{mathrsfs}
\usepackage{url}
\usepackage{natbib}

\newcommand{\mynorm}[2]{\| {#1} \|_{#2}}

\newcommand{\enorm}[1]{\mynorm{#1}{2}}
\newcommand{\znorm}[1]{\mynorm{#1}{0}}

\newcommand{\bigceil}[1]{\left\lceil{#1}\right\rceil}

\newcommand{\reals}{\mathbb{R}}

\renewcommand{\enorm}[1]{\mynorm{#1}{}}
\def\ninept{\def\baselinestretch{.95}\let\normalsize\small\normalsize}
\newtheorem{theorem}{Theorem}

\title{\large \bf A Trivial Observation related to Sparse Recovery}

\author{Suvrit Sra\\ Max-Planck Institute for Biological Cybernetics}
\begin{document}
\ninept
\maketitle
\begin{abstract}
  We make a trivial modification to the elegant analysis of Garg and Khandekar
  (\emph{Gradient Descent with Sparsification} ICML 2009) that replaces the
  standard Restricted Isometry Property (RIP), with another RIP-type property
  (which could be simpler than the RIP, but we am not sure; it could be as
  hard as the RIP to check, thereby rendering this little writeup totally
  worthless).
\end{abstract}
%
%
\section{Introduction}
Recently~\citet{gakh.icml09} presented an algorithm for solving
\begin{equation}
  \label{eq:3}
  \min_x\ f(x) = \enorm{y-\Phi x}^2\quad\text{s.t.}\quad\znorm{x} \leq s,
\end{equation}
which iteratively updates $x$ as
\begin{equation}
  \label{eq:8}
  x \gets H_s\bigl(x - 0.5\gamma^{-1}\nabla f(x)\bigr),
\end{equation}
where $H_s$ denotes the \emph{hard-thresholding} operator which sets all but
the $s$ largest (in magnitude) entries to zero. The most important part of
\citeauthor{gakh.icml09}'s paper is the elegant analysis which establishes
that under the restriction $\delta_{2s} < 1/3$ on the RIP constant of the matrix
$\Phi$, updating $x$ as per~\eqref{eq:8} solves~\eqref{eq:3} in
\emph{near-linear} time.

The Restricted Isometry Property or RIP (see~\citep{candes.rip} for a survey)
provides \emph{sufficient} conditions for sparse recovery. In particular,
define the \emph{isometry constant} of $\Phi$ as the smallest number $\delta_s$ such
that for all $s$-sparse signals $x \in \reals^n$
\begin{equation}
  \label{eq:9}
  (1-\delta_s)\enorm{x}^2 \leq \enorm{\Phi x}^2 \leq (1+\delta_s)\enorm{x}^2,
\end{equation}
where $\enorm{\cdot}$ denotes the standard $\ell_2$, i.e., Euclidean
norm. Under various restrictions on $\delta_{2s}$ different recovery guarantees
have been provided---see \citep{candes.rip} for example.

One can replace~\eqref{eq:9} by the potentially simpler property (or maybe
verifying this property is also as difficult as verifying the RIP). Let $\alpha_s >
0$ and $\beta_s$ be \textbf{some numbers} for which, for every $s$-sparse signal
$x$, the following version of the RIP holds,
\begin{equation}
  \label{eq:6}
  \alpha_s\enorm{x}^2 \leq \enorm{\Phi x}^2 \leq \beta_s\enorm{x}^2.
\end{equation}
Then, the elegant analysis of~\citet{gakh.icml09} (hereafter abbreviated as
GK) can be trivially changed to still hold for~\eqref{eq:6}. Theorem~1 shows
the noise-free case; a similar theorem for the noisy version can also be
derived, but is skipped for brevity.

\begin{theorem}[Sparse recovery]
  Let $\Phi$ be a measurement matrix for which $\beta_{2s} < 2\alpha_{2s}$ and let $x^*$
  be an $s$-sparse vector satisfying $y=\Phi x^*$. Then the \texttt{GraDes}
  algorithm of GK with $\gamma=\beta_{2s}$, computes an $s$-sparse vector $x \in
  \reals^n$ such that $f(x) \leq \epsilon$ in
  \begin{equation}
    \label{eq:5}
    \bigceil{\frac{\log(\enorm{y}^2/\epsilon)}{\log \alpha_{2s}/(\beta_{2s}-\alpha_{2s})}}\quad\text{iterations}.
  \end{equation}
\end{theorem}
The proof is a super-trivial modification to the proof of GK, and goes through
if one replaces $(1+\delta_{2s})$ by $\beta_{2s}$, and $(1-\delta_{2s})$ by $\alpha_{2s}$; Thus,
we omit further details to avoid verbosity. The interested reader can check
this fact in a few minutes herself by checking the nice paper of~\citep{gakh.icml09}.

The \textbf{only} point we wish to stress is that the RIP in its original form
is not critical to GK's proof, but rather just an RIP-type inequality of the
form~\eqref{eq:6}.

\bibliographystyle{plainnat}

\end{document}